\documentclass[prb,twocolumn,showpacs,amsmath,amssymb]{revtex4}

\usepackage{graphicx}% Include figure files
\usepackage{amssymb,amsmath}
\usepackage{dcolumn}% Align table columns on decimal point
\usepackage{bm}% bold math
\usepackage{color}

\begin{document}

\title{Coherent interactions between phonons and exciton or exciton-polariton condensates}

\author{D.V. Vishnevsky}
\affiliation{LASMEA, Nanostructure and Nanophotonics group, Clermont Universit\'{e}, Universit\'{e} Blaise Pascal, CNRS, 63177 Aubi\`{e}re Cedex France}

\author{N.A. Gippius}
\affiliation{LASMEA, Nanostructure and Nanophotonics group, 
Clermont Universit\'{e}, Universit\'{e} Blaise Pascal, CNRS, 63177 Aubi\`{e}re Cedex France}
\affiliation{A. M. Prokhorov General Physics Institute, RAS, Vavilova Street 38, Moscow 119991, Russia}

\author{I.A. Shelykh}
\affiliation{Physics Department, University of Iceland, Dunhagi-3,
IS-107, Reykjavik, Iceland}
\affiliation{International Institute of Physics, Av. Odilon Gomes de Lima, 1722, Capim Macio, CEP: 59078-400, Natal - RN, Brazil}
\affiliation{LASMEA, Nanostructure and Nanophotonics group, Clermont Universit\'{e}, Universit\'{e} Blaise Pascal, CNRS, 63177 Aubi\`{e}re Cedex France}

\author{D.D. Solnyshkov}
\affiliation{LASMEA, Nanostructure and Nanophotonics group, Clermont Universit\'{e}, Universit\'{e} Blaise Pascal, CNRS, 63177 Aubi\`{e}re Cedex France}

\author{G. Malpuech}
\affiliation{LASMEA, Nanostructure and Nanophotonics group, Clermont Universit\'{e}, Universit\'{e} Blaise Pascal, CNRS, 63177 Aubi\`{e}re Cedex France}

\begin{abstract}
We analyse the interaction of exciton and exciton-polariton condensates in semiconductor microcavity with a coherent acoustic wave. An analytical solution for the dispersion of excitations of coupled condensate-phonon system is found in the approximation of k-independent interactions. Accounting for k-dependence results in a stronger modification of the dispersion, and even in the appearance of the ''roton instability'' region.
\end{abstract}

\maketitle

\section{Introduction}

The interactions between different quasiparticles in the solid often lead to interesting effects. For example, exciton-polaritons or cavity polaritons appear in the microcavities \cite{KavokinBook} as the elementary excitations formed by the coupling of excitons and photons.
Having extremely small effective mass ($10^{-4}-10^{-5}$ of the effective mass of the free electrons), obeying bosonic statistics and efficiently interacting with each other, polaritons can undergo a transition into a collective phase
coherent state usually referred as polariton BEC \cite{Kasprzak2006}
accompanied by the onset of superfluidity\cite{MalpuechSSc}. Some indications of the occurence of this effect in real systems have been recently reported\cite{Amo2009,Amo2009a,Utsunomiya}. Another candidate for the achievement of the BEC in the domain of the condensed matter is a system of cold indirect excitons \cite{Butov,Butov1}.

Another type of elementary excitations in the solid are phonons
-- the collective oscillations of the crystal lattice that can also interact with light.
At relatively small light intensity emmition or absorption of phonons results in Mandelstam-Brillouin scattering with spectral energy shift on the phonon energy.
However with the increase of the intensity of the light the probability of the scattering process grows and it can become comparable with the lifetime of the scattered states. As it was shown in \cite{Ivanov1983,Keldysh1986a,Keldysh1986b} the strong reconstruction of the bulk phonon-polariton spectra occurs and the dispersions of both polaritons and phonons appear to depend strongly on the light intensity.

The interactions between the exciton-polaritons and strong acoustic waves has been a subject of recent studies \cite{Krizhanovskii} showing several interesting effects and opening many promising directions in the future, such as the possibility to observe polariton Bloch oscillations \cite{Flayac2011}.

In the present paper we consider the interaction between a condensate of quantum well (QW) excitons or cavity exciton-polaritons and a coherent phonon field, possessing the same dimensionality (2D or 1D). The coherent 2D acoustic wave can be obtained either by an external pumping (as in \cite{Krizhanovskii}) or by quantizing the phonon modes within an acoustic waveguide\cite{Jusserand2007} around the QW containing excitons (for cavity polaritons, the waveguide should be embedded inside the microcavity). If the phonons are not confined to the same dimensionality, they play a role of an incoherent reservoir providing exciton or exciton-polariton relaxation \cite{CavityPolaritons}, which is not the case we would like to consider here in order to obtain strong coupling.

In the phonon-photon or phonon-exciton coherent interaction models developed before\cite{Ivanov}, the phonon field has usually been treated as an external classical field. This approach is valid under strong external pumping, whereas the model we develop can be used both for the external pumping case and for the case when phonons are created by the excitons themselves.

\section{Formalism}

We treat our system consisting of three interacting classical fields: complex fields $\psi_\pm$ corresponding to right and left circular polarized excitons or exciton-polaritons and a real vector field of the lattice displacements $\textbf{u}$ corresponding to phonons. The energy of polariton or exciton system is measured from the minima of the dispersion curve. The Lagrangian of the system reads:

\begin{widetext}

\begin{eqnarray}
\mathcal{L}=\frac{i\hbar}{2}\sum_{s=\pm}\left(\psi_s\partial_t\psi_s^\ast-\psi_s^\ast\partial_t\psi_s\right)-\frac{\hbar^2}{2m}\sum_{s=\pm}\left(\nabla\psi_s
\right)\left(\nabla\psi_s^\ast\right)+\frac{1}{2}\left\{\rho\left(\partial_t\textbf{u}\right)^2-Y\left[\left(\partial_xu_x\right)^2+\left(\partial_yu_y\right)^2\right]\right\}-\\
\nonumber-\frac{\alpha_1}{2}\sum_{s=\pm}|\psi_s|^4-\alpha_2|\psi_+|^2|\psi_-|^2-g\textrm{div}(\textbf{u})\sum_{s=\pm}|\psi_s|^2
\end{eqnarray}

\end{widetext}
The first line corresponds to free polaritons described by means of their macroscopic wavefunctions $\psi_s$ and acoustic phonons described interms of the lattice displacement field $\bm{u}=(u_x,u_y)$. The second line to their mutual interactions: the term with $\alpha_1$ describes interactions of the polaritons of the same circular polarization, term with $\alpha_2$- interactions of the polaritons of the opposite circular polarization, term with $g$- polariton-coherent phonon interaction. For the moment, we neglect the wavevector dependence of this interaction and $g$ should be considered as a constant determined by the deformation potential. $m$ is an effective mass of cavity polaritons and $\rho$ and $Y$ are the density and the Young's modulus, determining the sound velocity $c_s=\sqrt{Y/\rho}$.

The equations of motion for the considered fields are Lagrange equations ($\psi_s$ and $\psi_s^\ast$ should be considered as independent functions):

\begin{eqnarray}
\frac{\partial}{\partial t}\left(\frac{\partial\mathcal{L}}{\partial(\partial_t\psi_s)}\right)+
\sum_{i=x,y}\frac{\partial}{\partial x_i}\left(\frac{\partial\mathcal{L}}{\partial(\partial_{x_i}\psi_s)}\right)-
\left(\frac{\partial\mathcal{L}}{\partial\psi_s}\right)=0\\
\frac{\partial}{\partial t}\left(\frac{\partial\mathcal{L}}{\partial(\partial_tu_i)}\right)+
\sum_{i=x,y}\frac{\partial}{\partial x_i}\left(\frac{\partial\mathcal{L}}{\partial(\partial_{x_i}u_i)}\right)-
\left(\frac{\partial\mathcal{L}}{\partial u_i}\right)=0
\end{eqnarray}

This gives:

\begin{eqnarray}
i\hbar\frac{\partial\psi_s}{\partial t}=-\frac{\hbar^2}{2m}\nabla^2\psi_s+\left(\alpha_1|\psi_s|^2+\alpha_2|\psi_s|^2\right)\psi_s+ \\
\nonumber +g\textrm{div}(\textbf{u})\psi_s\\
\rho\frac{\partial^2\textbf{u}}{\partial t^2}=Y\nabla^2\textbf{u}+g\nabla\left[\sum_{s=\pm}|\psi_s|^2\right]-\frac{\rho\gamma_\phi}{\hbar}\frac{\partial \phi}{\partial t}+P_{ph}(\textbf{r},t)
\end{eqnarray}
Taking the divergence of the second equation and introducing a scalar phonon field

\begin{equation}
\phi=\textrm{div}(\textbf{u})
\end{equation}
one gets

\begin{eqnarray}
i\hbar\frac{\partial\psi_s}{\partial t}=-\frac{\hbar^2}{2m}\nabla^2\psi_s+\left(\alpha_1|\psi_s|^2+\alpha_2|\psi_s|^2\right)\psi_s+g\phi\psi_s-\\
\nonumber-i\gamma_\psi\psi_s+P_\psi({\textbf{r},t})\\
\rho\frac{\partial^2\phi}{\partial t^2}=\nabla^2\left[Y\phi+g\sum_{s=\pm}|\psi_s|^2\right]-\frac{\rho\gamma_\phi}{\hbar}\frac{\partial \phi}{\partial t}+P_{ph}(\textbf{r},t)
\end{eqnarray}
where, in order to have the most general description, we have introduced the terms corresponding to the resonant pumping of the polariton and phonon modes ($P_\psi$ ans $P_\phi$ respectively) and the finite lifetime of the polaritons and phonons described by terms containing $\gamma_\psi$ and $\gamma_\phi$ respectively.

\section{Analytical solution}

In this section we calculate analytically the dispersion of excitations of coupled exciton or exciton-polariton and phonon modes with a k-independent interaction $g$. For simplicity we neglect here the spin of the polaritons, as our main goal here is to investigate the effects of spin- independent polariton- phonon coherent coupling. Then, one has the system of the equations for two coupled fields:

\begin{eqnarray}
i\hbar\frac{\partial\psi}{\partial t}=-i\gamma_\psi-\frac{\hbar^2}{2m}\nabla^2\psi+\alpha|\psi|^2\psi+g\phi\psi+P({\textbf{r},t})\label{GPPsiSpinless}\\
\rho\frac{\partial^2\phi}{\partial t^2}=\nabla^2\left[Y\phi+g|\psi|^2\right]-\frac{\rho\gamma_\phi}{\hbar}\frac{\partial \phi}{\partial t}+P_{ph}(\textbf{r},t) \label{GPPhiSpinless}
\end{eqnarray}

Now, consider a spatially homogeneous resonant pump of the polaritonic field under normal incidence, $P=P_0e^{-i\omega_0 t},P_{ph}=0$. Then, introducing $\psi=\Psi e^{-i\omega_0 t}$ one gets

\begin{eqnarray}
i\hbar\frac{\partial\Psi}{\partial t}=-(\hbar\omega_0+i\gamma_\psi)\Psi-\frac{\hbar^2}{2m}\nabla^2\Psi+ \\
\nonumber +\alpha|\Psi|^2\Psi+g\phi\Psi+P\label{GPPsiSpinless}\\
\rho\frac{\partial^2\phi}{\partial t^2}=\nabla^2\left[Y\phi+g|\Psi|^2\right]-\frac{\rho\gamma_\phi}{\hbar}\frac{\partial \phi}{\partial t} \label{GPPhiSpinless}
\end{eqnarray}
Looking for a spatially homogeneous stationary solution $\Psi_0,\phi_0$ one gets

\begin{eqnarray}
-\left(\omega_0+i\gamma_\psi\right)\Psi_0+\alpha|\Psi_0|^2\Psi+g\phi_0\Psi_0+P=0\\
Y\phi_0+g|\Psi_0|^2=F
\end{eqnarray}
where $F$ is hydrostatic pressure in the absence of the polaritons.
In what follows we assume it to be zero.

\begin{eqnarray}
\phi_0=-g|\Psi_0|^2/Y
\label{Phi0}
\end{eqnarray}
and for determination of the polariton field one has:

\begin{eqnarray}
-\left(\omega_0+i\gamma_\psi\right)\Psi_0+\left(\alpha-\frac{g^2}{Y}\right)|\Psi_0|^2\Psi_0+P=0
\label{ScalarBistability}
\end{eqnarray}

This is the same equation as for the case with polariton-polariton
 interactions only but with a renormalized polariton-polariton interaction constant,

\begin{eqnarray}
\tilde{\alpha}=\alpha-\frac{g^2}{Y}
\end{eqnarray}

One sees, that as correction to the interaction constant is negative, polariton-polariton interactions mediated by acoustic phonons are attractive. This is not surprising, as similar interactions for the electrons give attractive BCS potential.

The equation \ref{ScalarBistability} is well studied in context of quantum microcavities and can lead e.g. to the S-shaped dependence of the intensity of the polariton field on pump intensity, which results in a bistability and hysteresis \cite{Baas04, Gippius04e}.

Let us now calculate the dispersion of the weak elementary excitations. Following a standard procedure, one represents the solutions in a following form

\begin{eqnarray}
\Psi=\Psi_0+Ae^{i(\textbf{kr}-\omega t)}+B^\ast e^{-i(\textbf{kr}-\omega^\ast t)}\\
\phi=\phi_0+Ce^{i(\textbf{kr}-\omega t)}+C^\ast e^{-i(\textbf{kr}-\omega^\ast t)}
\end{eqnarray}
which should be then put into Eqs. \ref{GPPsiSpinless},\ref{GPPhiSpinless}. Considering deviations from equilibrium being small, one should then carry out the procedure of linearization, i.e. in the resulting equations keep only the terms linear in $A$,$B$,$C$.

The system of equations for determination of the small amplitudes $A$,$B$,$C$ reads:

\begin{widetext}
\begin{eqnarray}
\left(\begin{array}{ccc}-\hbar\omega-\chi & \alpha\Psi_0^2& g\Psi_0\\ \alpha\Psi_0^{\ast 2} & \hbar\omega-\chi & g\Psi_0^\ast\\-gk^2\Psi_0^\ast & -gk^2\Psi_0 & \rho\omega^2-Yk^2+i\frac{\rho\gamma_\phi}{\hbar}\end{array}\right)\left(\begin{array}{c}A\\B\\C\end{array}\right)=0
\end{eqnarray}

where $\chi=(\hbar\omega_0+i\gamma_\psi)-\left(2\alpha-\frac{g^2}{Y}\right)|\Psi_0|^2-\frac{\hbar^2k^2}{2m}$.

Thus the dispersions of the elementary excitations can be determined from the following equation:

\begin{eqnarray}
\left|\begin{array}{ccc}-\hbar\omega-\chi & \alpha\Psi_0^2& g\Psi_0\\ \alpha\Psi_0^{\ast 2} & \hbar\omega-\chi & g\Psi_0^\ast\\-gk^2\Psi_0^\ast & -gk^2\Psi_0 & \rho\omega^2-Yk^2+i\frac{\rho\gamma_\phi}{\hbar}\end{array}\right|=0
\end{eqnarray}
\end{widetext}

Here we used the relation \ref{Phi0} between the amplitudes of the excitonic and phonon fields.

\begin{figure}[h]
  \includegraphics[width=0.4\textwidth,clip]{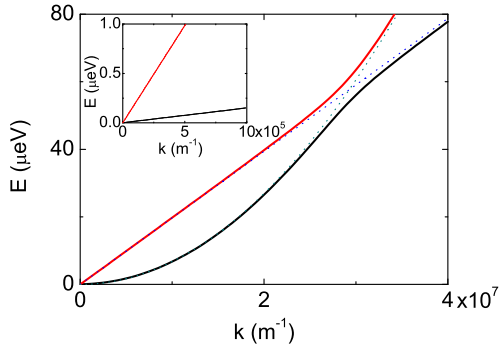}\\
  \caption{(Color online) Dispersion of elementary excitations of an exciton condensate strongly coupled with the phonon field. Dashed lines show the original excitations, solid lines show the renormalized dispersions. Inset: same dispertions for the small k-vectors}
  \label{Fig1}
\end{figure}

The consideration of the incoherent continous pump of the polariton system accompanied by the onset of the polariton BEC should be made in a different way. In this case, one can assume that in a stationary regime the finite lifetime of the polaritons is compensated by the pump, and calculations can be done in the same way as for particles with infinite lifetime with no external coherent pumping (i.e. one puts $P=0,\gamma_\phi=\gamma_\psi=\infty$). Besides, the energy of the macroscopically occupied mode is not pinned by the frequency of the resonant pump $\omega_0$ but is given by the chemical potential of the system  $\mu=\left(\alpha-\frac{g^2}{Y}\right)|\Psi_0|^2$.  In this case, the dispersions of the elementary excitations read

\begin{eqnarray}
\left|\begin{array}{ccc}-\hbar\omega+\beta & \alpha n& g\sqrt{n}\\ \alpha n & \hbar\omega+\beta & g\sqrt{n}\\-gk^2\sqrt{n} & -gk^2\sqrt{n} & \rho\omega^2-Yk^2\end{array}\right|=0
\end{eqnarray}
where $\beta=\alpha n+\frac{\hbar^2k^2}{2m}$.

This gives a following equation for $\omega(k)$, the dispersion of the new quasiparticles in the strongly-coupled system of bogolons (BEC elementary excitations)\cite{Yamamoto} and phonons:
%
% \begin{eqnarray}
% \left(\rho\omega^2-Yk^2\right)\left(E_0(k)^2+2\alpha nE_0(k)-(\hbar\omega)^2\right)+\\
% \nonumber+2g^2k^2nE_0(k)=0
%\end{eqnarray}
%
%or

\begin{eqnarray}
-(\hbar^2\rho)\omega^4+\left[\rho E_0(k)\left(E_0(k)+2\alpha n\right)+Y\hbar^2 k^2\right]\omega^2+\\
\nonumber+k^2E_0(k)\left[2g^2n-Y\left(E_0(k)+2\alpha n\right)\right]=0
\end{eqnarray}
with $ E_0(k) = \frac{\hbar^2k^2}{2m}$.
This equation gives rise to two solutions, close to the original dispersions in the low-$k$ limit and showing the typical anticrossing at higher wavevectors. As expected, in the region of small $k$ the dispersions are sound-like,

\begin{equation}
\omega_{1,2}=v_{1,2}k
\end{equation}

where

\begin{eqnarray}
v_{1,2}=
\sqrt{\frac{1}{2\rho}
\left[
\frac{\rho\alpha n}{m}+Y
\pm
\sqrt{\left(\frac{\rho\alpha n}{m}-Y\right)^2+\frac{4\rho g^2 n}{m}}
\right]
}
\end{eqnarray}

Figure 1 shows the dispersions calculated with the use of this equation (solid lines) in comparison with the original dispersions at zero exciton-phonon coupling $g=0$. The inset demonstrates the linear character of the dispersions near zero. Here and below we consider only exciton condensates, although the theory developed applies to both exciton and exciton-polaritons  (the latter in the parabolic approximation). The material parameters used are those of GaAs, as in \cite{Jusserand,Wertz}. Since the dispersion of phonons is not steep at all compared to that of polaritons, the anticrossing of the two dispersions takes place at relatively small wave vectors ($10^3~$m$^{-1}$). For excitons the situation is much more favorable, because their energy grows slower with $k$ than that of polaritons, and thus the anticrossing takes place at much larger $k$ ($10^7$ m$^{-1}$). In order to demonstrate the effects linked with the presence of the condensate we have to choose its density carefully, so that the speed of sound in the condensate $c=\sqrt{\alpha_1 n/m}$ lies below that of phonons, otherwise the branches will not anticross.

\section{Wavevector dependence}

\begin{figure}[h]
  \includegraphics[width=0.5\textwidth,clip]{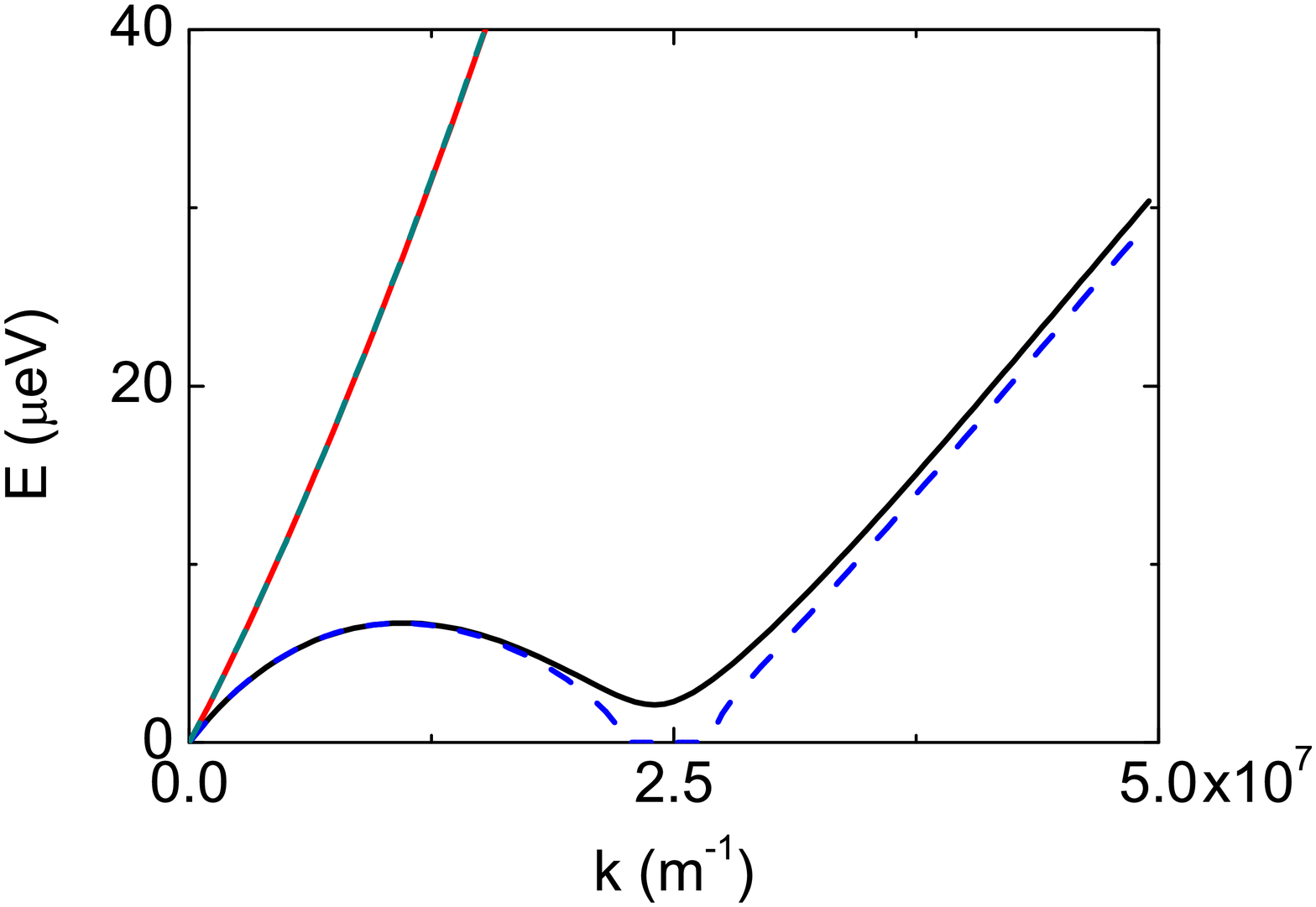}\\
  \caption{(Color online) Dispersion of the excitations in the exciton condensate coupled with waveguided phonon mode. The wavevector dependence of interactions is taken into account. Solid lines correspond to the stable situation with an additional valley appearing in the dispersion, whereas dashed lines show the unstable case with imaginary dispersion.}
  \label{Fig2}
\end{figure}

In order to obtain an analytical solution for the dispersions of excitations, we had to assume that the interactions between excitons and phonons were independent on the wavevector. However, in realistic systems this is not the case: $g(k)$ first increases with wavevector and then drops down to zero exponentially due to the overlap integrals between the exciton wavefunction and the amplitude of the acoustic wave. We will calculate this dependence for the structure placed in an acoustic waveguide.

The matrix element of exciton-acoustic phonon interaction is\cite{CavityPolaritons}:

\begin{equation}
M^{ac} (\vec q) = \sqrt {\frac{{\hbar q}}{{2\rho c_s SL}}} G(\vec q_{||} ,q_z )
\end{equation}

where $S$, $L$ - dimensions of the QW, $\vec q$ - acoustic wave vector. One can represent the function $G(\vec q_{||} ,q_z )$ as:

\begin{equation}
G(\vec q_{||} ,q_z ) = D_e I_e^{||} (\vec q_{||} )I_e^ \bot  (q_z ) + D_h I_h^{||} (\vec q_{||} )I_h^ \bot  (q_z )
\end{equation}

$D_e$, $D_h$ - are the deformation coefficients for electrons and holes and $I_e^ \bot  (q_z )$, $I_h^ \bot  (q_z )$, $I_e^ \bot  (\vec q_{||} )$, $I_h^ \bot  (\vec q_{||})$ are the overlap integrals between the phonon and exciton mode.

In order to calculate these integrals we have to obtain the distribution of the amplitude of the guided phonon wave inside the waveguide.
Taking into account that QW thickness is much less than the thickness of waveguide, we can assume that the amplitude of the acoustic field is constant over the QW in the $z$ direction. This value can be found from the waveguide equations\cite{Markuse}:

\begin{equation}
\phi_0  = \frac{1}{{2(\frac{{\sin \kappa d}}{\kappa } + \frac{{\cos \kappa d}}{{\sqrt {[n_1^2  + n_2^2 ]q_{||}^2  - \kappa ^2 } }})}}
\end{equation}

where $d$ is a thickness of the waveguide, $n_1$, $n_2$ are acoustic diffraction coefficients of waveguide layers, and $\kappa$ is a parameter which can be found by solving an equation:

\begin{equation}
\kappa \tan \kappa d = \sqrt {[n_1^2  + n_2^2 ]q_{||}^2  - \kappa ^2 }
\end{equation}

Putting the value of $\phi_0$ into the overlap integrals, we can obtain the final value of $g(q_{||})$:

\begin{widetext}
\begin{equation}
  g({q_{||}}) = {\phi _0}({q_{||}})\sqrt {\frac{{\hbar q}}{{2\rho {c_s}SL}}} \left( {{D_e}{{\left[ {1 + {{\left( {\frac{{{m_e}{q_{||}}a_b^{2D}}}{{2({m_e} + {m_h})}}} \right)}^2}} \right]}^{ - {\raise0.7ex\hbox{$3$} \!\mathord{\left/
 {\vphantom {3 2}}\right.\kern-\nulldelimiterspace}
\!\lower0.7ex\hbox{$2$}}}}} \right. +  \hfill
  \left. {{D_h}{{\left[ {1 + {{\left( {\frac{{{m_h}{q_{||}}a_b^{2D}}}{{2({m_e} + {m_h})}}} \right)}^2}} \right]}^{ - {\raise0.7ex\hbox{$3$} \!\mathord{\left/
 {\vphantom {3 2}}\right.\kern-\nulldelimiterspace}
\!\lower0.7ex\hbox{$2$}}}}} \right) \hfill \\
\end{equation}
\end{widetext}
First, we have calculated the dispersion of the excitations numerically, in order to compare the result with the analytical solution found above. The results of these calculations are shown in figure 2 (solid lines). One can see that if the maximum of the $k$-dependent exciton-phonon interactions coincides with the crossing of the original branches, the interactions can lead to a much larger splitting and even to the appearance of a valley on the dispersion, similar to the "roton minimum" \cite{LandauRotons, ShelykhRotons}. Dashed lines correspond to the extreme case, when the dispersion of the excitations becomes imaginary and the system becomes unstable against any small perturbation in this wavevector region.

We have next studied the behavior of the system characterized by such interactions between bogolons and phonons. An anticrossing between two quantum levels can give rise to so-called Rabi oscillations, which take place if only one of the original levels is excited initially. We demonstrate them in our system by solving numerically the equations for the exciton condensate order parameter and for the displacement field (\ref{GPPhiSpinless}), assuming a homogeneous condensate and a propagating acousting wave in the $x$ direction as the initial conditions. The system is homogeneous in the $y$ direction and thus the problem is reduced to 1D.

\begin{figure}[h]
  \includegraphics[width=0.5\textwidth]{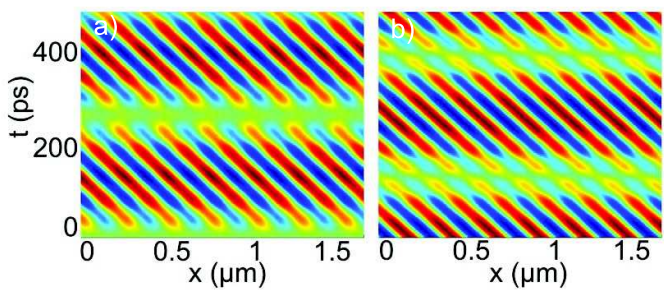}\\
  \caption{(Color online) Rabi oscillations between the excitations of the exciton condensate (bogolons) and phonons: the amplitude of the propagating waves in each component oscillates in time, as expected. Panel (a) shows the condensate density as a function of coordinate and time, panel (b) shows the amplitude of the acoustic wave.}
  \label{Fig3}
\end{figure}

\begin{figure}[h]
  \includegraphics[width=0.5\textwidth]{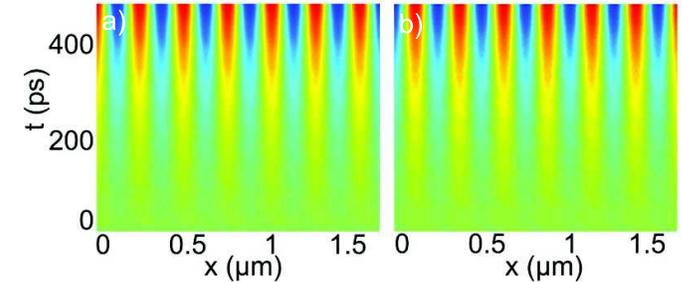}\\
  \caption{(Color online) Exponential growth of a weak excitation in the unstable case. Panel (a) shows the evolution of the condensate density over time, panel (b) shows the amplitude of the acoustic wave over time.}
  \label{Fig4}
\end{figure}

Figure 3 obtained this way demonstrates the Rabi oscillations between the two types of propagating waves: the bogolons (panel a) and the phonons (panel b). Initially, the exciton condensate is homogeneous (no bogolons), whereas the phonon field contains a single monochromatic propagating wave. This situation corresponds to an excited atom in an empty cavity in original Rabi oscillations. After 150 ps, the amplitude of the oscillations of the acoustic wave drops to zero, whereas the condensate density exhibits a running wave. This corresponds to the atom being in the ground state and a photon in the cavity mode, if one continues the analogy with Rabi oscillations. After about 300 ps, the acoustic wave becomes strong again while the bogolons disappear completely, which corresponds to one period of Rabi oscillations.

Finally, we have simulated the unstable situation corresponding to the dashed lines in figure 3. An excitation is created in the phonon field with its wavevector in the region with imaginary dispersion of excitations (flat real part). Figure 4 shows the results of our simulations: the small initial perturbation (a monochromatic wave) grows over time, but does not propagate in space, as expected from the flat real part and positive imaginary part of the dispersion (see fig. 3). The perturbations grow in both components (condensate and phonon field, panels (a) and (b)). This interesting effect cannot be described with the previous models (such as ref.\onlinecite{Ivanov}), because they do not allow to take into account the effect produced by excitons or photons on the acoustic field itself.

\section{Conclusions}

We have studied the interaction of QW excitons and cavity exciton-polaritons (condensed or uncondensed) with a coherent phonon field (guided acoustic mode). We have found an analytical solution and studied the renormalized dispersion of excitations in different situations, demonstrating the anticrossing of the two branches, an appearance of a valley at $k\ne 0$. If the lower dispersion branch touches 0, the system becomes unstable against any small perturbation in the corresponding wavevector range. We have performed numerical simulations to demonstrate the Rabi oscillations between bogolons and phonons in the stable regime and the exponential growth of the perturbations in the unstable regime.

The authors acknowledge the support of the joint CNRS-RFBR PICS project and FP7 ITN Spin-Optronics (237252) and IRSES ``POLAPHEN'' (246912) projects.

\end{document}